\documentstyle[12pt,here,psfig,doublespace]{article}

\begin{document}

{\tiny .}
\vspace{1.5cm}
\begin{center}

{\LARGE {\bf Molecular Gas in the Host Galaxy\\[3mm] of a Quasar at 
Redshift z=6.42 }}\\[0.5cm] 
{\bf Fabian Walter$^{1*}$, Frank Bertoldi$^+$, Chris Carilli$^*$, 
Pierre Cox$^{++}$, K.Y.~Lo$^*$, Roberto Neri$^{\&}$,
Xiaohui Fan$^{**}$, Alain Omont$^\#$, Michael A.~Strauss$^{\#\#}$,
Karl M.~Menten$^+$}
\end{center}

\noindent
$^*$ National Radio Astronomy Observatory, P.O. Box 0, Socorro, NM 87801,
USA \\
$^+$ Max--Planck--Institut f\"ur Radioastronomie, Auf dem H\"ugel 69,
53121 Bonn, Germany \\
$^{++}$ Institut d'Astrophysique Spatiale, Universite de Paris-Sud, 91405
Orsay, France \\
$^{\&}$ IRAM, 300 Rue de la Piscine, 38406 St--Martin--d'Heres, France\\
$^{**}$ Steward Observatory, University of Arizona, 933 N. Cherry Ave.,
Tucson, AZ 85721, USA \\
$^\#$ Institut d'Astrophysique de Paris, CNRS \& Universit\'e Paris 6, 75014 Paris, France\\
$^{\#\#}$ Princeton University Observatory, Princeton, NJ 08544, USA \\
$^1$ Jansky Fellow\\
\vspace{1cm}
Revised Nature manuscript: 2003-05-04710\\

\newpage

{\bf Observations of the molecular gas phase in quasar host galaxies
provide fundamental constraints on galaxy evolution at the highest
redshifts. Molecular gas is the material out of which stars form; it
can be traced by spectral line emission of carbon--monoxide (CO).  To
date, CO emission was detected in more than a dozen quasar host
galaxies with redshifts (z) larger 2, the record holder being at
z=4.69$^{1-3}$.  At these distances the CO lines are shifted to longer
wavelengths, enabling their observation with sensitive radio and
millimetre interferometers$^{4}$.  Here we present the discovery of CO
emission toward the quasar SDSS\,J114816.64+525150.3 (hereafter
J1148+5251) at a redshift of z=6.42$^{5,6}$ (when the universe was
only $\sim$1/16 of its present age).  This is the first detection of
molecular gas at the end of cosmic reionization. The presence of large
amounts of molecular gas (M(H$_2$)=2.2$\times10^{10}$\,M$_\odot$) in
an object at this time demonstrates that heavy element enriched
molecular gas can be generated rapidly in the earliest galaxies.}

The source J1148+5251 is a luminous quasar which is thought to be
powered by mass accretion onto a supermassive black hole of mass
1--5$\times10^9$ M$_{\odot}^{7}$. Optical spectra of J1148+5251 show a
clear Gunn--Peterson trough$^{5,6}$ (i.e. Ly $\alpha$ absorption by
the neutral intergalactic medium), indicating that this quasar is
situated at the end of the epoch of reionization$^{8}$.  Thermal
emission from warm dust was detected at millimetre wavelengths,
implying a far infrared (FIR) luminosity of $1.3\times
10^{13}$\,L$_\odot^{~9}$, corresponding to about 10$\%$ of the
bolometric luminosity of the system (we assume
H$_0\!=\!71$\,km\,s$^{-1}\,$Mpc$^{-1}$, $\Omega_{\Lambda}$=0.73 and
$\Omega_{m}$=0.27 throughout). We searched for molecular gas in
J1148+5251 using the National Radio Astronomy Observatory's (NRAO)
Very Large Array (VLA) and the IRAM Plateau de Bure interferometer
(PdBI) in February-April 2003. The VLA observations have higher
spatial resolution, whereas the PdBI has better spectral resolution.
We observed the CO(3-2), (6--5) and (7--6) lines which are shifted to
46.6, 93.2 and 108.7 GHz, respectively, at $z
\sim$6.42. For J1148+5251 the possible redshift range based on broad
optical emission lines is $z = 6.35^5$ to $6.43^{7}$. We searched this
entire range for CO(3--2) emission using the VLA at a resolution of 50
MHz (320 km s$^{-1}$, $\Delta$z=0.008) per channel. CO(3--2) emission
is clearly detected at high Signal to Noise in the channel centred at
46.6149 GHz (see Figs.~1 and 2), corresponding to a redshift of
z=6.418$\pm$0.004.  The PdBI covered a redshift range from
z=6.40--6.44 and the CO emission is also detected at high significance
(Fig.~2)$^{10}$. In this letter we concentrate on the VLA results; a
more detailed analysis of the entire dataset will be presented
elsewhere$^{10}$.

The velocity--integrated CO(3-2) flux is S$_{\rm CO(3-2)} \Delta
v=$0.18$\pm0.04$ Jy\,km\,s$^{-1}$. On the Kelvin scale
(1\,Jy\,=\,250\,K for a 1.5$''$ beam) the source has a peak line flux
of 0.14\,K and a line integral of 44$\pm10$\,K\,km\,s$^{-1}$.  The CO
luminosity is given by$^{11}$: $${\rm L}^{\prime}
=\!3.25\times10^7\times\,{\rm S}_{\rm CO}\Delta v\,[\rm
{Jy\,km\,s}^{-1}] \times \nu_{\rm obs}^{-2}\,[{\rm GHz}]\times {\rm
D}_{\rm L}^2\,[{\rm Mpc}] \times (1+z)^{-3}\, {\rm
K\,km\,s}^{-1}\,{\rm pc}^{2}, $$ where D$_{\rm L}$ is the luminosity
distance. For J1148+5251 we obtain: L$^{\prime}_{\rm
CO(3-2)}=2.7\times10^{10}$\,K\,km\,s$^{-1}$\,pc$^{2}$.  The total
molecular gas mass, M(H$_2$), can be derived from the relation
M(H$_2$) $=\!\alpha\times$L$^{\prime}_{\rm CO (1-0)}$. In Galactic
molecular clouds the conversion factor is $\alpha \sim 2.0$
M$_\odot$\,(K\,km\,s$^{-1}$\,pc$^{2})^{-1}$ $^{12}$, but here we use
the value $\alpha \sim 0.8$
M$_\odot$\,(K\,km\,s$^{-1}$\,pc$^{2})^{-1}$, appropriate for
ultraluminous infrared galaxies$^{13}$ (ULIRGs, L$_{\rm FIR} \ge
10^{12}$ L$_\odot$) and nuclear starburst galaxies$^{14}$. Assuming a
constant brightness temperature from CO(3--2) to CO(1--0)$^{10}$, this
leads to M(H$_2$)=2.2$\times10^{10}$\,M$_\odot$ for J1148+5251.

If the FIR luminosity of J1148+5251 is powered by star formation, the
implied star formation rate (SFR) is 3000\,M$_\odot$\,yr$^{-1}$
$^{9}$.  Comparing this to the molecular gas mass implies a short gas
depletion timescale of order $10^7$ years. However, given the high
luminosity of the quasar it is possible that some fraction of the dust
is heated by the quasar, in which case the star formation rate quoted
above becomes an upper limit (resulting in a longer gas depletion
timescale).  The FIR continuum--to--CO line ratio for J1148+5251 is
large, 440 L$_\odot$\,(K km s$^{-1}$ pc$^2)^{-1}$, which is an order of
magnitude larger than for normal star forming galaxies (5-50)$^{15}$,
but is comparable to that seen in ULIRGs and other high redshift
quasars$^{9}$. Some authors have suggested that high
continuum--to--line ratios indicate high star formation
efficiencies$^{15}$.  Alternatively, dust heating by the AGN could
lead to large continuum--to--line ratios.

A dynamical mass for the system can be derived from the CO
observations, assuming that the CO is rotationally supported.  The
upper limit to the CO source diameter derived from Gaussian fitting is
1.5$''$ (1$''$=5.46\,kpc). A lower limit to the source diameter can be
derived from the measured brightness temperature (T$_{\rm obs}$) by
assuming an intrinsic brightness temperature T$_{\rm b}$:
$\Omega_S/\Omega_B$=(T$_{\rm obs}$/T$_{\rm b}$)(1+z), where $\Omega_S$
and $\Omega_B$ are the source and beam solid angles. The most extreme
case is a single optically thick emission region, in which case
T$_{\rm b}$ equals T$_{\rm ex}$--T$_{\rm CMB}$, the excitation
brightness temperture minus the continuum temperature of the cosmic
background at z=6.4 ($\sim20$\,K).  This gives a minimum source
diameter of 0.2$''$(T$_{\rm b}/50\,$K)$^{-1/2}$. Assuming a rotational
velocity of v$_{\rm rot}=130 \sin^{-1}(i)$\,km\,s$^{-1}$ (Fig.~2),
where $i$ is the inclination angle with respect to the sky plane,
yields a range for the dynamical mass of M$_{\rm dyn}=2 ~ {\rm to} ~
16 \times10^{9}\times\sin^{-2}(i)$ M$_{\odot}$.  This is comparable to
the molecular gas mass in J1148+5251, implying that either molecular
gas dominates the dynamics of the system, or that the plane of
rotation is close to the sky plane.  The optical spectrum of the
quasar shows no significant reddening$^5$, despite the large dust mass
of the host galaxy$^9$. This suggests that the disk of molecular gas
(and dust) is oriented close to the sky plane such that the
line--of--sight to the nucleus is relatively unobscured.  We note that
the mass of the central black hole$^{4}$ likely constitutes a
significant fraction ($\sim1-10$\%) of the total dynamical mass in the
inner few kpc of J1148+5251, in contrast to what is generally found
for nearby galaxies$^{16,17}$.

The CO redshift of z=6.419 (Fig.~2) corresponds to the systemic
redshift of the host galaxy of J1148+5251, since it traces the
extended molecular gas distribution in the quasar and is not
associated with emission supposedly emerging from energetic processes,
e.g. lines of shocked outflow gas or gas related to AGN accretion.
The latter has been found to be shifted significantly in frequency
with respect to the systemic redshift$^{18}$. Indeed, studies of high
ionization UV lines (in particular SiIV)$^{5}$ in J1148+5251 yield a
redshift of z=6.37 corresponding to a velocity offset of
$\sim$2000\,km\,s$^{-1}$.  The CO redshift is however in good
agreement with the redshift derived from the low ionization MgII
line$^{7}$.

Measuring an accurate redshift is particularly important to determine
the state of the IGM around the quasar (the `proximity effect'). In
the optical spectrum$^{5,7}$ there is essentially zero emission
present in the wavelength range corresponding to Ly$\alpha$ at
z=5.7--6.33 (due to the Gunn--Peterson effect); emission for $z>6.33$
can be attributed to the ionized medium around the quasar. Using the
source redshift of 6.419 results in a Str\"omgren sphere around
J1148+5251 with a comoving radius of R$_S$=4.8\,Mpc. An estimate for
the age for this sphere (and hence for the quasar itself) can be
derived from R$_S$ using$^{19,6}$ $$\dot N_{ph}/(10^{58}\,{\rm
s}^{-1})\times t_q/(10^7\,{\rm yr})=0.34\times[R_S/(4.8\,{\rm
Mpc})\times(1+z_q)/7.419)]^3$$ and assuming $\dot N_{\rm
ph}=0.2-1.3\times10^{58}\,{\rm s}^{-1}$ $^{20,21,6}$. This calculation
results in an age of order $10^7$\,yr for the quasar activity in
J1148+5251. Interestingly, this timescale is comparable to the
formation timescale for the central black hole, which has an e-folding
timescale for accretion of order $4\times10^7$ yr (assuming a
radiative efficiency of 0.1 and that the black hole is accreting at
the Eddington limit)$^{7}$, suggesting that the AGN ionizes a
significant volume around the quasar.

The fact that we detect CO emission implies that the process of
enrichment of the ISM in J1148+5251 with heavy elements is relatively
advanced, i.e., that significant star formation has occurred in the
quasar host galaxy prior to $z\!=\!6.4$. This conclusion is supported
by the fact that both strong metal emission lines$^{5}$, and thermal
emission from warm dust$^{9}$, were detected from J1148+5251. A recent
optical study of a z=6.28 quasar even suggests supersolar
metallicities in these early objects$^{22}$. Assuming a Galactic
abundance for J1148+5251 we derive an order of magnitude estimate for
the total mass in CO of
M(CO)$\sim\!3\times10^7$\,M$_\odot~~^{10}$. This amount of C and O
could be produced relatively quickly by $\sim\!10^7$ hundred solar
mass Population III stars$^{23}$ with lifetimes
$<10^{7}$\,yr. Likewise, if enrichment was achieved by conventional
supernovae$^{24}$ and assuming a star formation rate of order
3000\,M$_\odot$\,yr$^{-1}$ $^{~9}$, we estimate a time for enrichment
of order $10^7$\,yr. We consider this time a lower limit for
enrichment of the ISM since redistribution and cooling of the gas will
occur on timescales of order $10^8$\,yr. This implies that the ISM
enrichment process in J1148+5251 presumably started at redshifts
$z\!>\!8$. This is in agreement with the recent results from the
Wilkinson Microwave Anisotropy Probe (WMAP) which indicate that the
first onset of star formation most likely occured at redshifts
z$\sim15-17$ ($\sim250$\,Myr after the Big Bang)$^{25,26,27}$.

The presence of Ly$\alpha$ emitting galaxies and luminous quasars at
the end of cosmic reionization ($z\!>\!6.3$), at a time when the IGM
was at least $1\%$ neutral, was clearly demonstrated$^{5,28}$. This
epoch of reionization represents a key bench--mark in cosmic structure
formation, indicating the formation of the first luminous structures.
Detecting a large reservoir of molecular gas in this epoch
demonstrates the existence of the requisite fuel for active star
formation in primeval galaxies. The existence of such reservoirs of
molecular gas at early times implies that studies of the youngest
galaxies will be possible at millimetre and centimetre wavelengths,
unhindered by obscuration by the neutral IGM.\\

{\bf References}

1. Ohta, K. et al. Detection of molecular gas in the quasar BR1202-0725 at
redshift z=4.69, {\em Nature} {\bf 382}, 426--428 (1996)

2. Omont, A. et al. Molecular gas and dust around a radio-quiet quasar
at redshift 4.69, {\em Nature} {\bf 382}, 428--431 (1996)

3. Carilli, C.L. et al.\ High-Resolution Imaging of Molecular Line
Emission from High-Redshift QSOs, {\em Astron. J.} {\bf 123}, 1838--1846
(2002)

4. Carilli, C.L. \& Blain, A.W. Centimeter searches for molecular
line emission from high-redshift galaxies, {\em Astrophs. J.} {\bf
569}, 605--610 (2002)

5. Fan, X. et al. A survey of $z>5.7$ quasars in the sloan digital sky
survey. II. Discovery of three additional quasars at $z>6$, {\em
Astron. J.} {\bf 125}, 1649--1659 (2003)

6. White, R.L., Becker, R.H., Fan, X. \& Strauss, M.A. Probing the
ionization state of the universe at z$>$6, {\em Astron. J.}  in press
(2003)

7. Willott, C.J., McLure, R.J. \& Jarvis, M.J. A $3\times10^9$
M$_\odot$ black hole in the quasar SDSS J1148+5251 at z=6.41, {\em
Astrophys. J.} {\bf 587}, L15--L18 (2003)

8. Loeb, A. \& Barkana, R. The Reionization of the Universe by the
First Stars and Quasars, {\em Ann. Rev. Astron. Astroph.} {\bf 39}, 19--66
(2000)

9. Bertoldi, F., Carilli, C.L., Cox, P., Fan, X., Strauss, M.A.,
Beelen, A., Omont, A. \& Zylka, R. Dust emission from the most distant
quasars, {\em Astron. Astrophys. Letter} in press (2003)

10. Bertoldi, F. et al., {\em Astron. Astroph. Letters}, 2003, subm.

11. Solomon, P.M., Radford, S.J.E. \& Downes, D. Molecular gas content
of the primeval galaxy IRAS 10214+4724, {\em Nature} {\bf 356}, 318--321
(1992)

12. Strong, A.W., et al. Diffuse continuum gamma rays from the Galaxy
observed by COMPTEL, {\em Astron. Astroph.}, {\bf 292}, 82--91 (1994)

13. Downes D. \& Solomon, P. Rotating Nuclear Rings and Extreme
Starbursts in Ultraluminous Galaxies, {\em Astroph. J.} {\bf 507},
615--654 (1998)

14. Weiss, A., Neininger, N., Huttemeister, S. \& Klein, U. The effect
of violent star formation on the state of the molecular gas in M82,
{\em Astron. Astrophys.} {\bf 365}, 571--587 (2001)

15. Solomon, P.M., Downes, D., Radford, S.J.E. \& Barrett, J.W. The
Molecular Interstellar Medium in Ultraluminous Infrared Galaxies, {\em
Astroph. J.} {\bf 478}, 144 (1997)

16. Ferrarese, L. \& Merritt, D. A fundamental relation between
supermassive black holes and their host galaxies, {\em Astroph. J.}
{\bf 539}, L9--L12 (2000)

17. Gebhardt, K. et al. A relationship between the nuclear black hole
mass and galaxy velocity dispersion, {\em Astroph. J.} {\bf 539},
L13--16 (2000)

18. Richards, G.T. et al. Broad emission-line shifts in quasars: an
orientation measure for radio-quiet quasars? {\em Astron. J.} {\bf
124}, 1--17 (2002)

19. Haiman, Z. \& Cen, R. A constraint on the gravitational lensing
magnification and the age of the redshift z=6.28 quasar SDSS
1030+0524, {\em Astroph. J.} {\bf 578}, 702--707 (2001)

20. Elvis, M. et al. Atlas of quasar energy distributions, {\em
Astroph. J. Sup.}  {\bf 95}, 1--68 (1995)

21. Telfer, R.C., Zheng, W., Kriss, G.A. \& Davidsen, A.F. The
Rest--Frame Extreme--Ultraviolet Spectral Properties of Quasi-stellar
Objects, {\em Astroph. J.} {\bf 565}, 773--785 (2002)

22. Pentericci, L. et al.\ VLT Optical and Near-Infrared Observations
of the z = 6.28 Quasar SDSS J1030+0524, {\em Astron. J.} {\bf 123},
2151--5158

23. Heger, A. \& Woosley, S.E. The nucleosynthetic signature of
population III, {\em Astroph. J.} {\bf 567}, 532--543 (2002)

24. Arnett, A. Massive star evolution and SN 1987A, {\em Astroph. J.} 
{\bf 383}, 295--307 (1991)

25. Cen, R. Implications of WMAP Observations On the Population III
Star Formation Processes, {\em Astroph. J. Letters} subm.

26. Kogut et al. Wilkinson Microwave Anisotropy Probe (WMAP) First
Year Observations: TE Polarization, {\em Astroph. J.}, subm. (2003)

27. Spergel et al. First Year Wilkinson Microwave Anisotropy Probe
(WMAP) Observations: Determination of Cosmological Parameters, {\em
Astroph. J.}, subm. (2003)

28. Hu, E.M. et al. A redshift z=6.56 galaxy behind the cluster Abell
370, {\em Astroph. J.} {\bf 568}, L75--L79 (2002)

\newpage

{\bf Acknowledgments:} The VLA is operated by the National Radio
Astronomy Observatory (NRAO), a facility of the National Science
Foundation (NSF), operated under cooperative agreement by Associated
Universities, Inc. (AUI). This work is partly based on observations
carried out with the IRAM Plateau de Bure Interferometer. IRAM is
supported by INSU/CNRS (France), MPG (Germany) and IGN (Spain).\\

{\bf Competing Interest Statement:} The authors declare that they have
no competing financial interests.\\

{\bf Correspondence} should be addressed to F.W. (e-mail:
fwalter@nrao.edu)\\

\newpage

\begin{figure}[B]
   \begin{center} \hspace{-1.5cm} 
   \psfig{figure=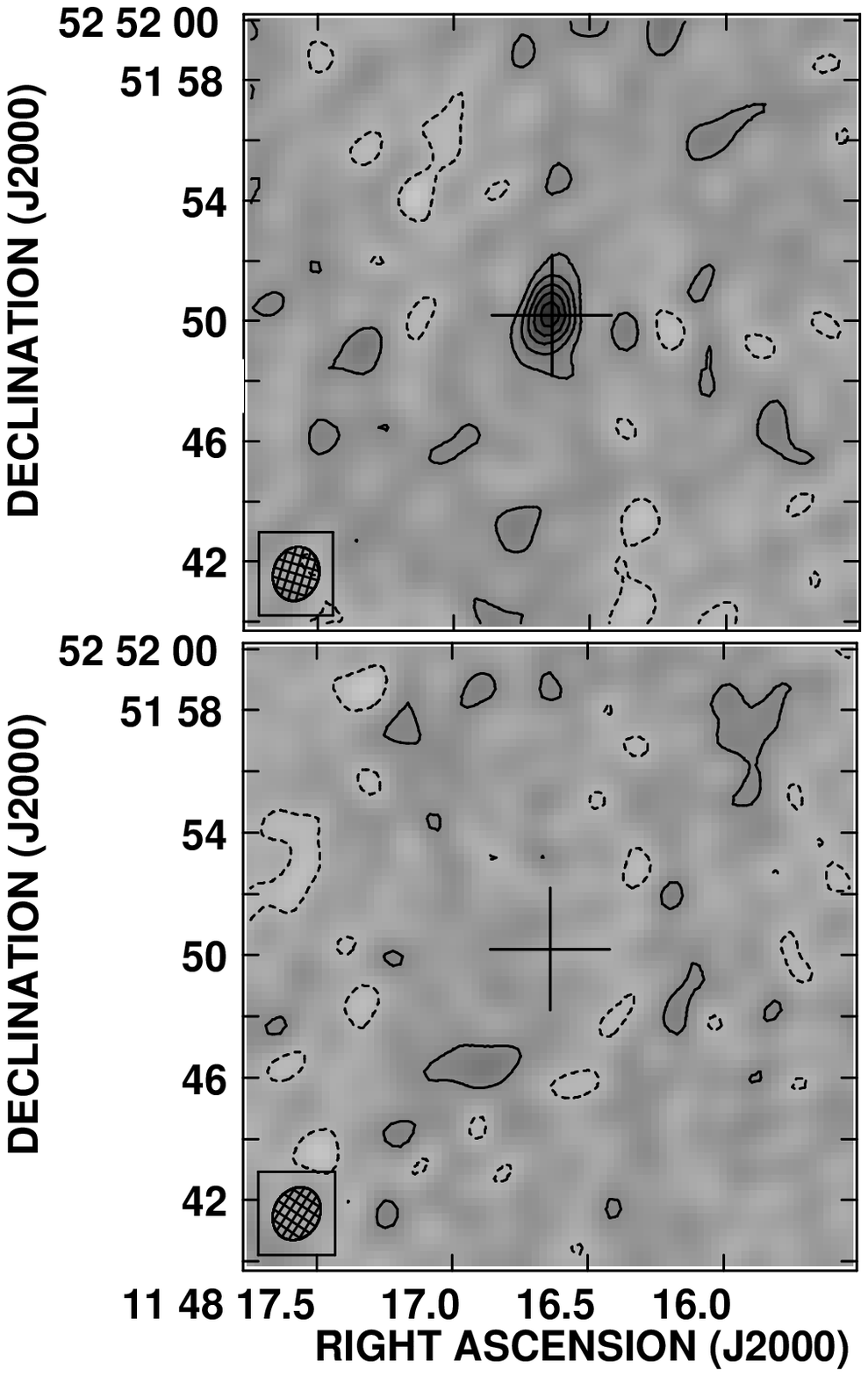,width=130mm,clip=t,angle=0}
   \end{center} \vspace{0cm}

\end{figure}

{\em Fig. 1: CO detection in J1148+5251.} {\em Top:} VLA CO(3--2)
emission detected at 46.6149\,GHz (z=6.418, bandwith:~50\,MHz,
$\Delta$z=0.008). {\em Bottom:} Co--added VLA emission in the
neighboring line--free channels. The (1\,$\sigma$) noise in both
images is 0.057\,mJy\,beam$^{-1}$ and contours are plotted at -0.1,
0.1, 0.2, 0.3, 0.4 and 0.5\,mJy\,beam$^{-1}$. The peak flux in the top
panel is 0.570\,mJy\,beam$^{-1}$; i.e. J1148+5251 is detected at
10$\sigma$.  Based on the line free channels we derive a 2$\sigma$
upper limit to the continuum emission at 46.6 GHz (restframe
wavelength=870$\mu$m) of 0.10 mJy.  The cross indicates the optical
position of J1148+5251. The optical and radio positions are coincident
within the uncertainties ($\sim 0.1'' = 0.6$\,kpc). The observations
were taken with the VLA in the most compact configuration (D array,
maximum baseline: $\sim\!1\,$km, leading to a resolution of
$1.8''\!\times\!1.5''$; the beam is plotted in the lower left of each
panel). Only two 50\,MHz channels can be observed at once with the
VLA, i.e., the source has been observed repeatedly with different
frequency settings; observations were made using two polarizations and
2 IFs of 50 MHz each, scanning in frequency from 47.065 GHz to 46.515
GHz, corresponding to a CO(3--2) redshift range of 6.347 to 6.434.
The observing time for the channels shown are $\sim$20 hours,
respectively.  Observations were obtained in fast switching mode, and
the phase stability was excellent at all times. We used the nearby
quasar 11534+49311 (flux density = 1.6 Jy at 46.6\,GHz) for phase and
amplitude calibration. The absolute flux calibration was derived by
observing 3C286.  No evidence for strong gravitational lensing is seen
in optical images$^{5}$, rendering strong magnification
unlikely. However, the presence of an intervening galaxy at z=4.9 has
been suggested based on optical spectroscopy$^{6}$, and counts of
radio and optical sources in the vicinity of J1148+5251 argue for a
foreground cluster, such that weak magnification ($\sim$ factor two)
is possible$^{9,5}$.

\newpage

\begin{figure}[B]
\vspace{-5cm}
   \begin{center} \hspace{0cm} \hspace{2.5cm}
   \psfig{figure=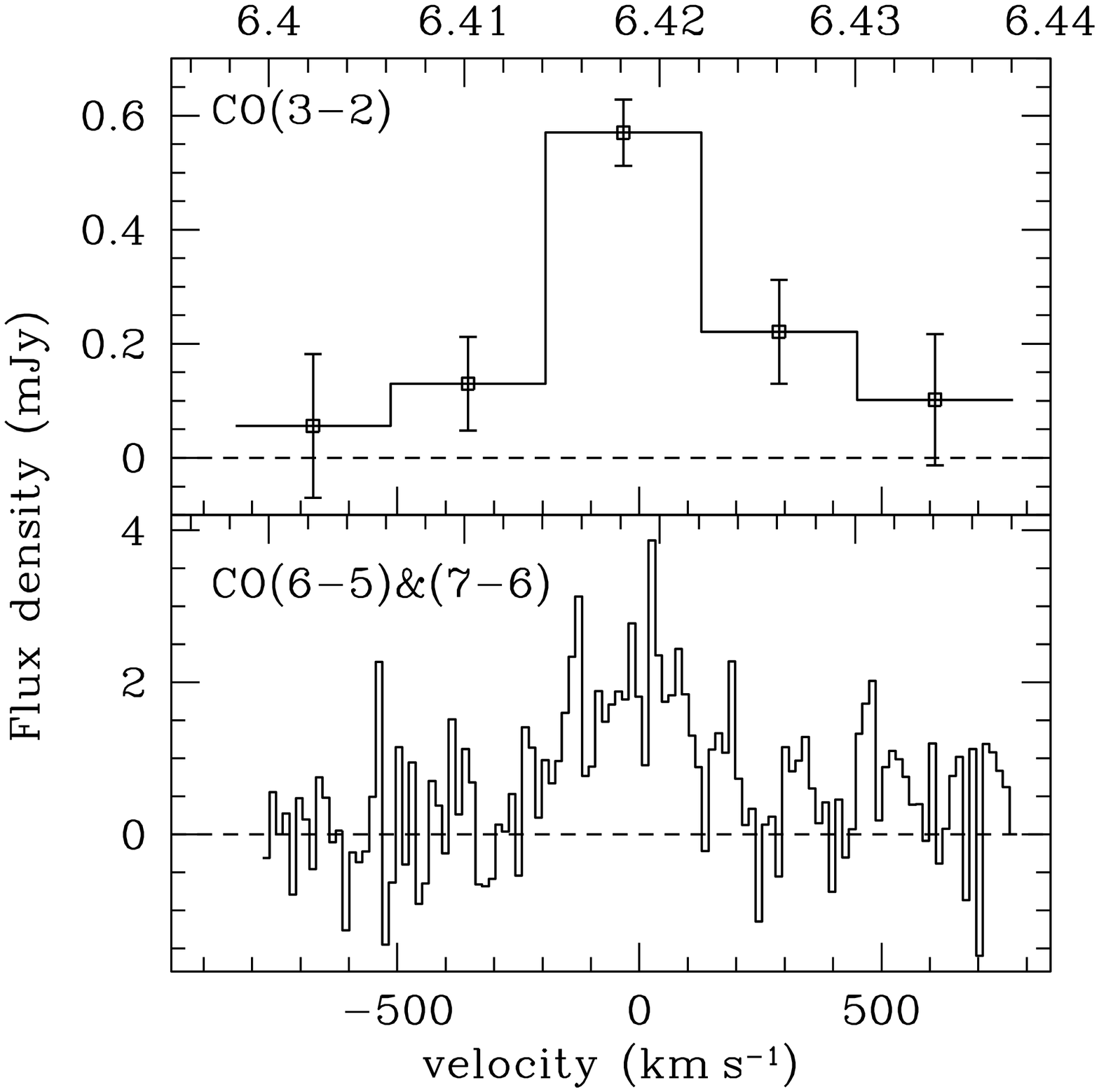,width=130mm,clip=t,angle=0}
   \end{center} \vspace{-0.8cm}

\end{figure}

{\em Fig. 2: The CO spectrum of J1148+5251.} The spectrum of the
CO(3--2) line of J1148+5251 at 320 km s$^{-1}$ resolution as observed
with the VLA (top panel). The plotted errors are $\pm1\sigma$. The
averaged CO(6--5) and CO(7--6) observations from the PdBI are shown in
the bottom panel$^{10}$ (channel width: 5\,MHz\,=\,13.8\,km\,s$^{-1}$,
noise per channel: 0.8\,mJy, beam size: $\sim5''$) to demonstrate the
consistency of the results obtained by the two instruments. A Gauss
fit to the PdBI data gives a velocity width of 250\,km\,s$^{-1}$ and a
redshift of z=6.419 (corresponding to `0' velocity).

\end{document}